\documentclass[10pt,letterpaper]{article}
\usepackage{opex3}

\begin{document}

\title{Macroscopic entanglement between a Bose Einstein condensate and a superconducting loop}

\author{Mandip Singh}

\address{Centre for Atom Optics and Ultrafast Spectroscopy, ARC
Centre of Excellence for Quantum-Atom Optics, Swinburne University
of Technology, Melbourne, Australia \\}

\email{naturemandip@gmail.com} 

\begin{abstract}
We theoretically study macroscopic entanglement between a magnetically trapped Bose-Einstein condensate and a superconducting loop. We treat the superconducting loop in a quantum superposition of two different flux states coupling with the magnetic trap to generate macroscopic entanglement. The scheme also provides a platform to investigate interferometry with an entangled Bose Einstein condensate and to explore physics at the quantum-classical interface.
\end{abstract}



\section{Introduction}
Entanglement is considered to be one of the most fundamental features of quantum mechanics. In addition, it is of great importance in the context of quantum information and quantum computation. In recent years, there have been considerable efforts to generate and preserve entanglement for quantum information processing. In particular, entanglement among macroscopic observables is of prime interest in order to explore how a system behaves at the interface of classical and quantum mechanics. Macroscopically entangled states are also promising candidates for the practical realization of a quantum computer.

In recent years there has been ground breaking progress in the field of manipulation of BECs. Nowadays, it is relatively easy to produce a BEC in a micro-magnetic trap on an atom chip \cite{han01}. A BEC in such traps can be coherently manipulated with RF fields \cite{sch05, hoff06} and microwaves \cite{tre06, tre04}. Since neutral atoms can be positioned a few microns from the chip surface and moved with nanometre resolution, atom chips provide a convenient platform to study the interaction between a BEC and a nearby surface \cite{jon03} including the study of fundamental quantum effects such as the Casimir-Polder interactions \cite{har05}. On the other hand the field of superconducting circuits is progressing rapidly in terms of technological implementation and realization of quantum coherent control of superconducting qubits \cite{mak01, moo99}. A macroscopic superposition of different magnetic flux states has been demonstrated \cite{fri00, van00} and quantum coherent dynamics of flux qubits have been realised \cite{chi03}. In addition, there are proposals to study the interaction between a superconducting circuit and a micromechanical device, in particular to entangle a micromechanical resonator with a Cooper pair box \cite{arm02}. Since a nanomechanical cantilever can be easily integrated on an atom chip a magnetically trapped Bose-Einstein condensate can be coherently coupled to the oscillation modes of a magnetized cantilever \cite{tre07}. Two spatially separated Rydberg atoms can also be capacitively coupled to each other through a thin superconducting wire \cite{andr04}. Atom chips made of superconducting substrate and wires carrying persistent current to manipulate trapped atoms have been reported \cite{nirr06, muk07}. The magnetic field exclusion due to Meissner effect in a superconducting wire can significantly modify the magnetic trap properties \cite{cano08}. Recently, there is a great interest to study the coupling between ultracold gases and solid state quantum systems to explore their quantum properties. We theoretically study a macroscopic entanglement between a superconducting loop and a BEC.  By exploiting the quantum mechanical properties of superconducting circuits and their interactions with nearby ultracold atoms or a BEC a macroscopic entanglement can be produced which provides a pathway to explore physics at the quantum-classical interface.
\begin{center}
\begin{figure}
\begin{center}
\includegraphics[scale=0.55]{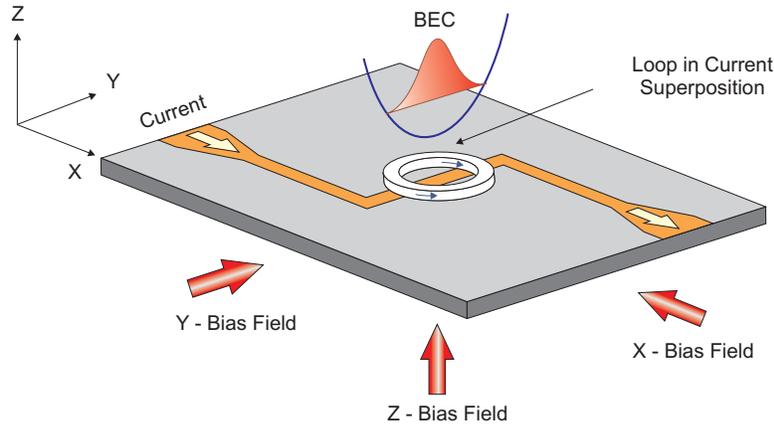}
\caption{\label{fig:chip} Schematic showing a superconducting loop situated symmetrically above a Z-shaped wire magnetic trap. The persistent current in the loop can exist in a quantum superposition in the clockwise and anti-clockwise directions. The magnetic field produced by the persistent current can perturb the BEC trapping potential where the degree of perturbation can be increased by moving the trap center closer to the loop. The superconducting loop can be biased appropriately by applying an external magnetic field along the z-direction.}
\end{center}
\end{figure}
\end{center}

\section{Superconducting loop on an atom chip}

 According to the present scheme, a superconducting loop existing in a quantum superposition of macroscopically distinct flux states can perturb the BEC trapping potential and under certain conditions a macroscopic entanglement can be generated. The macroscopic variables involved in the entangled state are the flux linked to the loop, the chemical potential of the BEC and the position and shape of the BEC wavefunction. The physical arrangement of the superconducting loop on the atom chip is shown in Fig~\ref{fig:chip}. Ultracold neutral atoms or a BEC can be trapped in a magnetic trap above the surface by applying an external bias field in the plane of a Z-shaped current-carrying wire. A superconducting loop is positioned symmetrically above the Z-wire and below the trap in such a way that there is zero net flux linked to it. The trap position and trap frequencies depend on the external bias fields and the current in the Z-wire. The magnetic field produced by a nearby superconducting loop carrying persistent current can perturb the magnetic trap potential through magnetostatic interactions. The sign and amplitude of the perturbation depends on the direction of the persistent current flowing in the loop and its position with respect to the trap centre.

A superconducting loop when placed in an external flux permits only discrete values of the net flux threaded through it, which is an integral multiple of the flux quantum. In other words, the super-current in the loop responds automatically to any change in the externally applied flux in order to keep the closed loop phase acquired by the wave function an integral multiple of 2$\pi$ . In the case of a superconducting ring interrupted by a Josephson tunnel junction the total energy (the sum of the magnetic energy of the loop and Josephson coupling energy of the junction) corresponds to a double-well potential of a flux variable threaded through the loop \cite{fri00}, where the left (right) well corresponds to persistent current flowing in a clockwise (anticlockwise) direction in the loop. The double well becomes symmetric when the applied flux is equal to half of the flux quantum. The barrier height between the wells depends on the critical current of the Josephson tunnel junction. The inter-well tunnelling can be controlled externally by replacing the single Josephson junction with split junctions (DC-SQUID) \cite{fri00}.

At low temperature the superconducting  loop can be prepared in a quantum superposition of two persistent current states flowing clockwise and anticlockwise by biasing it at half of the flux quantum. A BEC can be prepared in a Z-shaped wire magnetic trap far above the superconducting loop where the perturbation in the trapping potential caused by the loop is negligibly small. If the magnetic trap containing a BEC is moved adiabatically towards the superconducting loop carrying a persistent current the trapped BEC will adiabatically follow the perturbation in the potential induced by the magnetic field from the loop. However, if the loop is initially prepared in a quantum superposition of persistent current flowing clockwise and anti-clockwise it will perturb the magnetic trap potential in two different perturbations which can be distinguished by increasing the interaction with the loop leading to a macroscopic entanglement between the persistent current state of the loop and the state of the BEC in different perturbed potentials in the magnetic trap. One perturbed configuration of a BEC can differ from the other one in terms of its spatial distribution and the chemical potential. The signature of entanglement between the superconducting loop and the BEC can be observed through the atomic distribution in time of flight measurements.

\section{Entanglement between the loop and a BEC}

In order to study the macroscopic entanglement between the loop and the BEC we start by deriving the Hamiltonian of the superconducting loop coupled to the magnetic trap containing a BEC. In the case when the loop is biased at half of the flux quantum the double well governing the dynamics of the loop is symmetric. Therefore, at low temperature, the Hamiltonian of the superconducting loop can be treated as a two-level system.

\begin{equation}
\label{eq:1} H_{S}=E_{0}|0\rangle\langle0|+E_{0}|1\rangle\langle1|+J|1\rangle\langle0|+J|0\rangle\langle1|
\end{equation}
where $|0\rangle$ and $|1\rangle$ represent the ground state of the left and the right well, respectively, and $J$ is the tunnelling amplitude between them.

The Hamiltonian of an atom of mass $m$ in the ground state of the trap coupled to the superconducting loop in the case $J = 0$ can be written as

\begin{equation}
\label{eq:2}
 H_{T}= \int\hat{\Psi}^{\dag}(r){\Bigg[}\left(\frac{-\hbar^2}{2m}\nabla^2+ V(r)\right)\hat{I}+\Delta V_{0}(r,t)|0\rangle\langle0|
      +\Delta V_{1}(r,t)|1\rangle\langle1| {\Bigg]}\hat{\Psi}(r) \mathrm{d}r
\end{equation}

where $\Delta V_{0}(r,t)$ and $\Delta V_{1}(r,t)$ are the perturbations in the trap potential $V(r)$ due to coupling to the loop in states $|0\rangle$, $|1\rangle$, respectively, and $|0\rangle \langle0| + |1\rangle \langle1| = \hat{I} $. For a magnetic trap potential $V(r)= m_{F}g_{F}\mu_{B}B(r)$ the perturbation in $V(r)$ originates from the magnetic field produced by the persistent current in the loop which  affects the harmonic trapping field profile $B(r)$ of the Z-wire magnetic trap near its trap centre. In the case when loop is prepared in state $|0\rangle$ and the perturbation is increased adiabatically the trapped BEC follows the ground state $\phi_{0}(r,t)$ of the trapping potential $V(r)+ \Delta V_{0}(r,t)$ and likewise if the loop is initially prepared in state $|1\rangle$ the BEC follows the ground state $\phi_{1}(r,t)$ of the potential $V(r)+ \Delta V_{1}(r,t)$.
  However, under the adiabatic condition and for the loop in quantum superposition (of $|0\rangle$ and $|1\rangle$) the trapped BEC follows two different configurations $\phi_{0}(r,t)$ and $\phi_{1}(r,t)$ because the state of perturbation is coupled to the state of the loop. Therefore, the field operator $\hat{\Psi}(r,t)$ can be expanded as a linear combination of $\phi_{0}(r,t)|0\rangle\langle0|$ and $\phi_{1}(r,t)|1\rangle\langle1|$;

\begin{equation}
\label{eq:3} \hat{\Psi}(r,t)=\hat{a}_{0}\phi_{0}(r,t)|0\rangle\langle0|+\hat{a}_{1}\phi_{1}(r,t)|1\rangle\langle1|
\end{equation}

where $\hat{a}_{0}$ and $\hat{a}_{1}$   are the corresponding bosonic annihilation operators.
Therefore, from Eqs.~\ref{eq:1},~\ref{eq:2} and~\ref{eq:3} the Hamiltonian of the system can be written as

\begin{equation}
\label{eq:4} H(t)=E_{0}|0\rangle\langle0|+E_{0}|1\rangle\langle1|+\mu_{0}(t)\hat{a}^\dag_{0}\hat{a_{0}}|0\rangle\langle0|+ \mu_{1}(t)\hat{a}^{\dag}_{1}\hat{a}_{1}|1\rangle\langle1|
\end{equation}

where

\begin{equation}
\label{eq:5} \mu_{0}(t)=\int\phi^\dag_{0}(r,t)\left(\frac{-\hbar^2}{2m}\nabla^2+V(r)+\Delta V_{0}(r,t)\right)\phi_{0}(r,t) \mathrm{d}r
\end{equation}

and

\begin{equation}
\label{eq:6} \mu_{1}(t)=\int\phi^\dag_{1}(r,t)\left(\frac{-\hbar^2}{2m}\nabla^2+V(r)+\Delta V_{1}(r,t)\right)\phi_{1}(r,t) \mathrm{d}r
\end{equation}

are the energy eigenvalues in the case of two perturbed situations of the trap.

 Let us see how entanglement can be generated. In the first step a BEC of $N$ atoms is prepared in a Z-wire magnetic trap far away from the superconducting loop so that perturbation caused by the loop is negligibly small. In the second step the superconducting loop is prepared in a symmetric superposition $|S\rangle=(|0\rangle + |1\rangle)/\sqrt{2}$  and then the tunnelling amplitude $J$  is reduced to zero. In the third step the BEC is slowly brought closer to the superconducting loop (by increasing the $x$-bias magnetic field or by decreasing the current in the Z-wire) so that the perturbation to the magnetic trap is increased adiabatically and the BEC will follow two different configurations of the trap. As the two different perturbations grow in quantum superposition a macroscopic entanglement between the state of the BEC in the trap and the state of the superconducting loops established. At one point where the coupling is sufficiently strong two perturbed configurations of the trap will be different and distinguishable.

Therefore at $t=0$ the initial state of the system is $|\Psi,t=0\rangle=(|0\rangle+|1\rangle)|N,\phi(r,t=0)\rangle/\sqrt{2}$,
where $|N,\phi(r,t=0)\rangle$ is the state corresponding to $N$ atoms in the ground state $\phi(r,t)$ of the trap potential $V(r)$ in the case of no coupling with the flux loop. When the coupling is increased adiabatically the state of the system evolves under the unitary evolution $U(t) = \exp( -\frac{i}{\hbar}\int_0^{t}H (t')\mathrm{d}t')$ to

\begin{equation}
\label{eq:7} |\Psi,t\rangle  = \frac{e^{i\gamma_{0}(t)-i(E_{0}t+N\int_0^{t}\mu_{0}(t') \mathrm{d}t')/\hbar}}{\sqrt{2}}
  \times \left[|0\rangle|N,\phi_{0}(r,t)\rangle+e^{i\Phi(t)}|1\rangle|N,\phi_{1}(r,t)\rangle\right]
\end{equation}

where

\begin{equation}
\label{eq:8} \Phi(t)=N \int_0^{t}\frac{\mu_{0}(t')-\mu_{1}(t')}{\hbar} \mathrm{d}t'+\gamma_{1}(t)-\gamma_{0}(t)
\end{equation}

    and $|N,\phi_{0}(r,t)\rangle$ and $|N,\phi_{1}(r,t)\rangle$ are states corresponding to $N$ atoms in the ground states $\phi_{0}(r,t)$ and $\phi_{1}(r,t)$, respectively. In addition $\gamma_{0} (t)$ and $\gamma_{1}(t)$ represent the geometrical phase. Equation.~\ref{eq:7} is derived  using the following transformations implied by the adiabatic condition

   \begin{equation}
\label{eq:9a} e^{ -\frac{i}{\hbar}\int_0^{t}\mu_{0}(t')\hat{a}^\dag_{0}\hat{a_{0}}|0\rangle\langle0|\mathrm{d}t'}|0\rangle|N,\phi(r,t=0)\rangle = e^{ -\frac{iN}{\hbar}\int_0^{t}\mu_{0}(t')\mathrm{d}t'}e^{ i\gamma_{0}(t)}|0\rangle|N,\phi_{0}(r,t)\rangle
\end{equation}

  \begin{equation}
\label{eq:9b} e^{ -\frac{i}{\hbar}\int_0^{t}\mu_{1}(t')\hat{a}^\dag_{1}\hat{a_{1}}|1\rangle\langle1|\mathrm{d}t'}|1\rangle|N,\phi(r,t=0)\rangle
= e^{ -\frac{iN}{\hbar}\int_0^{t}\mu_{1}(t')\mathrm{d}t'}e^{ i\gamma_{1}(t)}|1\rangle|N,\phi_{1}(r,t)\rangle
\end{equation}

It is evident that $|\Psi,t\rangle $ (Eq .~\ref{eq:7}) represents a macroscopic entangled state between the states of the loop and the BEC in two different configurations. As  $\Phi(t)$ evolves $|\Psi,t\rangle $ spans different macroscopically entangled states.

\begin{center}
\begin{figure}
\begin{center}
\includegraphics[scale=0.50]{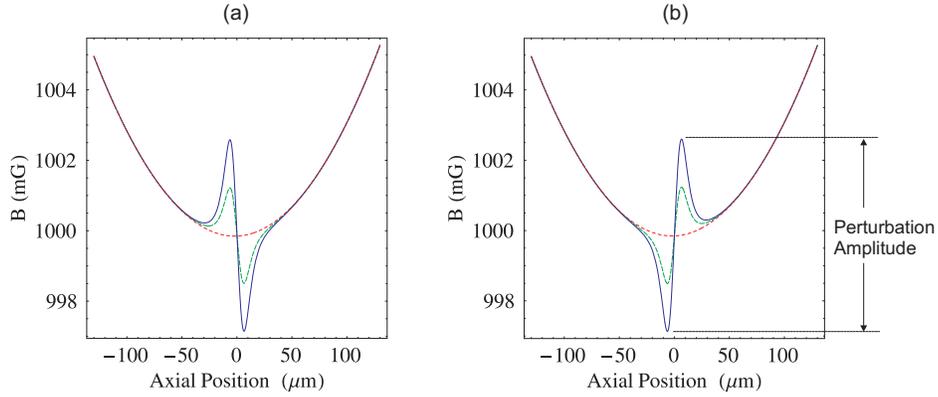}
\caption{\label{fig:clock} Magnetic field intensity profile along the axial direction of a Z-wire magnetic trap coupled to a superconducting loop when the persistent current flows (a) clockwise and (b) anti-clockwise. The dotted curve (red) represents the harmonic field profile without any coupling. The dashed curve (green) and solid line (blue) show the effect of perturbation when a persistent current in the loop generates a flux through the loop of magnitude one-quarter and one-half of a flux quantum, respectively. .}
\end{center}
\end{figure}
\end{center}

\begin{center}
\begin{figure}
\begin{center}
\includegraphics[scale=0.4]{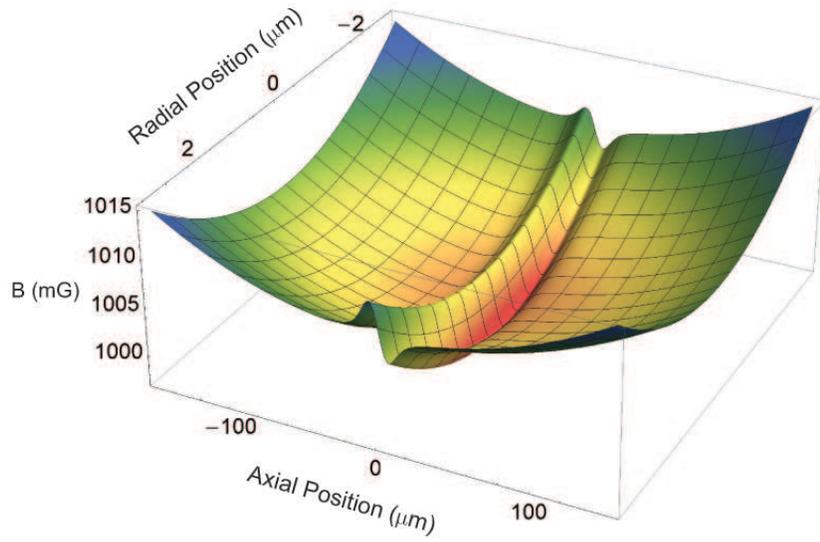}
\caption{\label{fig:threed} Magnetic field intensity profile around the trap minimum when a clockwise flow of persistent current generates a flux through the loop of magnitude one-half of a flux quantum. Radial and axial axes lie in the horizontal plane passing through the trap minimum.}
\end{center}
\end{figure}
\end{center}

\begin{center}
\begin{figure}
\begin{center}
\includegraphics[scale=0.4]{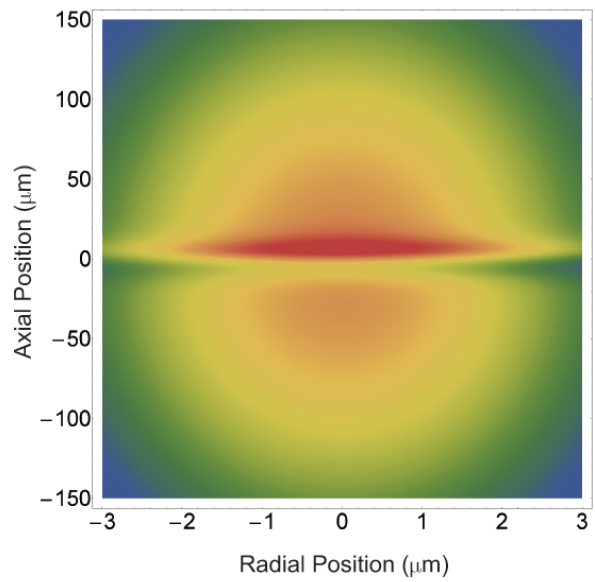}
\caption{\label{fig:densxy} Density plot showing the magnetic field intensity in the horizontal plane. The colour scheme has one to one correspondence with that indicated in Fig.~\ref{fig:threed}.}
\end{center}
\end{figure}
\end{center}

\begin{center}
\begin{figure}
\begin{center}
\includegraphics[scale=0.85]{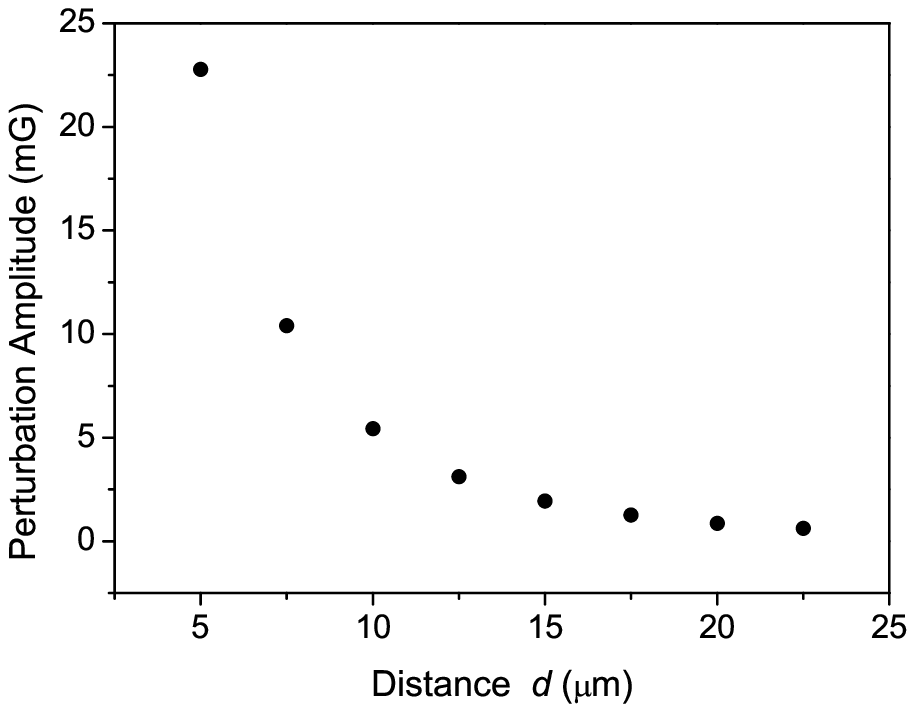}
\caption{\label{fig:distance} Perturbation amplitude as a function of distance ($d$) between the superconducting loop and the centre of the magnetic trap when the persistent current generates a flux through the loop of magnitude one-half of a flux quantum.}
\end{center}
\end{figure}
\end{center}

 In order to calculate the perturbation in the magnetic field intensity profile of the Z-trap a circular superconducting loop of diameter 10 $\mu$m is considered. The loop is positioned at a distance 10~$\mu$m below the trap minimum with its axis parallel to the $z$-axis. In order to bias the loop to produce a symmetric energy double well an applied field should generate a flux of magnitude one-half of a flux quantum through the loop. For the loop parameters given above a uniform magnetic field of about 0.1 G is required in the $z$-direction to bias the loop. The trap parameters are calculated for a Z-wire magnetic trap carrying 5 A DC current and 20 G uniform trapping bias magnetic field in the $x$-direction. The trap bottom is tuned to 1 G by applying an additional uniform bias field along the $y$-direction.  The perturbation in the magnetic field intensity profile of a Z-wire trap in the axial direction due to the magnetic field from the superconducting loop is shown in Fig.~\ref{fig:clock} for clockwise (a) and anticlockwise (b) flow of persistent current (same in magnitude). The dotted curve (red) represents the unperturbed harmonic profile when the persistent current is zero. The dashed line (green) and solid line (blue) represent the magnetic field intensity corresponding to the persistent current generating a flux through the loop of magnitude one-quarter and one-half of a flux quantum, respectively. A three-dimensional profile of the magnetic field intensity (around the trap centre) in a horizontal plane parallel to the $x$-$y$ plane and passing through the trap minimum is shown in Fig.~\ref{fig:threed}. The axial and radial axes refer to the weakly and strongly confining directions, respectively, in the horizontal plane. The corresponding density plot is shown in Fig.~\ref{fig:densxy}. The amplitude of the perturbation which is defined as the difference of the magnetic field intensity at the local maximum and the local minimum is about 5.5 mG for a persistent current generating a flux of magnitude one-half of a flux quantum through the loop as indicated in Fig.~\ref{fig:clock} (b). The chemical potential for $N$ atoms of $^{87}$Rb (for $|F = 2, m_{F} = 2\rangle$ with radial and axial trap frequencies 540~Hz and 10~Hz, respectively) in the absence of any perturbation is $0.01959 N^{2/5}$ mG. The amplitude of the perturbation should be greater than the chemical potential of the BEC in order to distinguish between the perturbed states. The distance between the minima of the two perturbed magnetic field profiles is of the order of the diameter of the superconducting loop. The strength of the perturbation depends on the distance of the magnetic trap centre from the superconducting loop and increases as the trap centre is displaced towards the loop. Figure~\ref{fig:distance} shows the perturbation amplitude as a function of distance ($d$) between the trap centre and the loop measured along the axis of the loop. It is evident from Fig.~\ref{fig:distance} that the perturbation amplitude at $d > 20~\mu m$ is negligibility small and increases rapidly as $d$ is reduced.

\section{Entangled atom interferometry and detection}

  The interference property of a Bose Einstein condensate is an important tool to study the coherence among the condensates. In an analogue of a double slit experiment a single BEC can be coherently split by gradually changing the harmonic confinement into a double well and an interference pattern can be observed by overlapping two split BEC wavefunctions after they are released \cite{shin04,sch05}. In such experiments BECs are not path entangled. Now
    it is important question: if we release the BEC from  macroscopically entangled state (Eq.~\ref{eq:7}) can we get an interference and how can we detect macroscopic entanglement between the superconducting loop and the BEC.?

   In order to make the  idea clear let us consider a single atom in the BEC $(N=1)$. The expectation value of the atomic density after the release is evaluated as

\begin{equation}
\label{eq:9} n(r,t) = \langle\Psi,t|\hat{\psi^{\dag}}(r) \hat{\psi}(r)|\Psi,t\rangle = \frac{1}{{2}}(|\phi_{0}(r,t)|^{2}+|\phi_{1}(r,t)|^{2})
\end{equation}

 This is an incoherent mixture with no interference; the interference term is suppressed because the which-path information about the BEC state is available through the entanglement with the superconducting loop. It is possible to know which trap configuration the atom occupies just by measuring the state of the loop. Therefore, in order to get interference the which-path information should be erased. This is possible if $|0\rangle$ and $|1\rangle$ are indistinguishable. On the other hand, in a new basis where $|+\rangle=\frac{1}{\sqrt{2}}(|0\rangle+|1\rangle)$, $|-\rangle=\frac{1}{\sqrt{2}}(|0\rangle-|1\rangle)$, $|S\rangle = \frac{1}{\sqrt{2}}(|N=1,\phi_{0}(r,t)\rangle+|N=1,\phi_{1}(r,t)\rangle)$ and $|A\rangle = \frac{1}{\sqrt{2}}(|N=1,\phi_{0}(r,t)\rangle-|N=1,\phi_{1}(r,t)\rangle)$ the state Eq.~\ref{eq:7} for $\Phi(t)=0$  can be written as (overall phase ignored)

\begin{equation}
\label{eq:10} |\Psi,t\rangle=\frac{1}{\sqrt{2}}[|+\rangle |S\rangle + |-\rangle |A\rangle]
\end{equation}
         It is evident from the state Eq.~\ref{eq:10} that when the superconducting loop is found in state $|+\rangle$ in repeated measurements the corresponding atomic distribution shows an interference term $\cos(2\pi r/\Lambda)$, where $2\pi/\Lambda = tmd/(\hbar(t^2+(m\sigma^2_{0}/\hbar)^2))$ (initial gaussian width $\sigma_{0}$ located at $\pm d/2$) \cite{shin04}.
          If the loop is found in $|-\rangle$ the atomic distribution exhibits an anti-interference term. Therefore, if the state of the superconducting loop is not measured then there will be no way to differentiate between interference and anti-interference and no interference pattern will emerge. In the case of more than one atom in the state Eq.~\ref{eq:7} the situation is different. If we treat the BEC state as the control qubit acting on the superconducting state and perform a CNOT transformation ($U_{CNOT}|N,\phi_{0}(r,t)\rangle |0\rangle = |N,\phi_{0}(r,t)\rangle |0\rangle$ and $U_{CNOT}|N,\phi_{1}(r,t)\rangle |1\rangle = |N,\phi_{1}(r,t)\rangle |0\rangle$) after releasing the BEC then the state Eq.~\ref{eq:7} transforms to

 \begin{eqnarray}
\label{eq:11}
        |\Psi,t\rangle & =& \frac{1}{\sqrt{2}}|0\rangle (|N,\phi_{0}(r,t)\rangle+e^{i\Phi(t)}|N,\phi_{1}(r,t)\rangle)
\end{eqnarray}

               There is no entanglement between the BEC and the superconducting loop now and in this case the centre of mass density of the BEC when released from the state Eq.~\ref{eq:11} should exhibit a pattern with $\Lambda^{'} = \Lambda/N$ \cite{wal04,mit04,mol99, sac00,bach04} in repeated time of flight measurements; this is a signature of the macroscopic entanglement Eq.~\ref{eq:7}. However, only those measurements which count exactly the same $N$ should be considered; therefore, a precise control for state preparation (Eq.~\ref{eq:7}) is required, where $\Phi(t)$ should not vary by more than over $\pi/2N$ from measurement to measurement \cite{bach04}. However, it might be difficult at present to implement such a macroscopic CNOT operation in order to get the state of interest Eq.~\ref{eq:11}.
               There is another way to observe a signature of the macroscopic entanglement between the loop and the BEC. If we measure the distribution of the BEC when it is released from the trap and correlate it with the state of the loop then it is evident from Eq.~\ref{eq:7} that whenever the superconducting loop is measured in $|0\rangle$ the measured particle density distribution should be $N|\phi_{0} (r,t)|^{2}$ and when it is measured in $|1\rangle$ the particle density distribution is given as $N|\phi_{1} (r,t)|^{2}$. This can be evidence that the superconducting loop is interacting with the BEC in the trap. However, if the superconducting loop is measured in state $|+\rangle$ then in this basis for the state Eq.~\ref{eq:7} the particle density distribution should shift either from $N|\phi_{0} (r,t)|^{2}$ or from $N|\phi_{1} (r,t)|^{2}$ \cite{bach04}. This shift in density distribution can be attributed to the existence of macroscopic entanglement of a pure condensate with a superconducting loop.

\section{Decoherence}
In context of a practical realization it is important to mention that the generation of macroscopic entanglement demands the validity of adiabatic turn-on of the perturbation potential. Therefore, the decoherence time for the loop must be larger than the time required to complete the adiabatic process so that the required entanglement can be observed. There are various factors \cite{van03} which can limit the decoherence time of a flux loop. Noise in the magnetic field, interaction of the loop with the state measuring apparatus and impurities in the substrate on which the loop is fabricated can decohere a flux superposition. At present the decoherence and relaxation times for a flux qubit of the order of 1-10 $\mu$s have been reported \cite{you05}. In order to satisfy the adiabatic condition while increasing the coupling between the superconducting loop and the magnetic trap the turn on process should be slow enough to satisfy $\frac{d\omega}{dt}\ll\omega^{2}$ where $\omega$ is the angular trap frequency. If change in the trap frequency during the perturbation turn-on process is of the order of the trap frequency then the time for the adiabatic approximation to hold can be estimated roughly from the trap frequencies. Since the effect of perturbation is much stronger along the axial direction (weaker confinement). Therefore, the inverse of axial trap frequency can give a rough estimation of the adiabatic time limit. For a typical axial trap frequency of 10~Hz the adiabatic time limit estimation is about 100~ms which requires a considerable improvement in the decoherence time of the flux loop. However, an accurate calculation of the time limit for the validity of the adiabatic approximation can be calculated by numerically solving the dynamics of BEC in the time dependent perturbation by minimizing the transition probability to other excited states from the instantaneous ground state of the potential.

In order to evaluate the time limit for the adiabatic approximation to hold it is important to know the functional form of the magnetic trap potential moving towards ($z$-direction) the superconducting loop which can be expressed as
 \begin{eqnarray}
\label{eq:12}
        V(x,y,z,t)&=& \frac{1}{2} m \omega_{x}^{2}(t)x^{2}+\frac{1}{2}m \omega_{y}^{2}(t)y^{2}+\frac{1}{2}m \omega_{z}^{2}(t)(z+z_{0}(t))^{2}+ V_{0}+\nonumber\\
        &&\frac{2 a(t) y}{\sigma^{2}}e^{-\frac{y^{2}}{\sigma^{2}}}|0\rangle\langle0|-\frac{2 a(t) y}{\sigma^{2}}e^{-\frac{y^{2}}{\sigma^{2}}}|1\rangle\langle1|
        \end{eqnarray}
where $\omega_{x}$, $\omega_{z}$ correspond to the radial trap frequencies and $\omega_{y}$ is the axial trap frequency and it is assumed that the state dependent perturbation is well defined. Since the trap considered here is predominantly anisotropic ($\omega_{x}, \omega_{z} >> \omega_{y}$) the effect of perturbation induced by the loop is more significant along the axial direction. The term $\frac{2 a(t) y}{\sigma^{2}}e^{-\frac{y^{2}}{\sigma^{2}}}$ represents the mean perturbation along the axial direction of the magnetic trap when the loop is in $|1\rangle$ state. When the loop exists in state $|0\rangle$, the sign of the perturbation is changed. The displacement of the trap towards the loop and how fast the perturbation is induced depends on $z_{0}(t)$. The trap frequency in the case of a Z-wire trap is inversely proportional to the distance of the trap centre from the Z-wire and the offset field at the trap centre. A variation in the trap frequency when the trap is moved towards the loop can be avoided by synchronously varying the offset field. According to the numerical calculations shown in Fig.~\ref{fig:distance} the perturbation amplitude is negligibly small for $d> 20~\mu$m and grows rapidly as $d$ is reduced. Therefore a BEC can be prepared at about $d \geq 20~\mu m$ and by decreasing $d$ to about $ 10~\mu$m a perturbation amplitude of about 5.5 mG can be obtained. The function $B_{0} + k_{0}y^{2}+\frac{2 a y}{\sigma_{0}^{2}}e^{-\frac{y^{2}}{\sigma_{0}^{2}}}$ is fitted to the axial magnetic field profile shown in  Fig.~\ref{fig:clock} (a) for clockwise flow of supercurrent generating a flux equal to half of a flux quantum (solid blue line) where values of the fit parameters are $B_{0}= 999.85$~mG, $k_{0}=0.00031$~mG/$\mu$m$^{2}$, $\sigma_{0} = 10.13$ $\mu$m and $a=-32.0$~mG$\mu$m. Therefore, the potential profile can be obtained by using $V(r)=m_{F}g_{F}\mu_{B}B(r)$. The sign of $a$ changes for the anti-clockwise flow of persistent current. One can use the potential described in Eq.~\ref{eq:12} in order to study the dynamics of BEC coupling to the superconducting loop and to numerically evaluate the time limit for the adiabatic approximation.

\section{Remarks}
In the context of experimental realization it is important to consider that the magnetic field from the Z-wire and the bias field should be less than the critical field of the superconducting loop. The experiment can be constructed by utilizing flip-chip technology, where the Z-wire and the superconducting loop can be constructed on two different substrates which can be bonded together with high precision with an appropriate gap between them. The Z-wire can also be constructed from a superconducting material in order to reduce the technical noise in the current. The atom chip can be shielded from the background radiation by a gold coated copper shield. The whole assembly can be mounted on a cold finger with the chip pointing up side down. The ultracold atoms can be prepared in a different chamber and magnetically transported to radiation shielded chamber containing the chip where they can be trapped and evaporatively cooled \cite{muk07} down to a BEC.

There are various factors that can destroy a macroscopic entangled state described by Eq.~\ref{eq:7}. Loss of a single atom from this state can easily destroy the entanglement by sharing its information with the environment. However, a superconducting flux superposition itself is prone to environment-induced decoherence \cite{mak01}. In order to satisfy the adiabatic condition while increasing the coupling between the superconducting loop and the magnetic trap the coupling turn-on process should be slow enough but should be completed before the decoherence time limit of the superconducting loop otherwise excitations can occur in the BEC.

In conclusion we have shown how the coherent dynamics of a superconducting loop can be used to generate a macroscopic entanglement of a BEC on an atom chip. Such a macroscopic entanglement could be useful to explore fundamental quantum mechanics by studying how quantum mechanical effects behave at the macroscopic level and how they decohere with the system parameters. It may also be possible to explore decoherence between the superconducting circuit and the BEC by varying the trap parameters and the size of the system such as the number of atoms in the BEC. It has been shown how to realize interference of an entangled BEC and in addition how it can be used to detect a signature of macroscopic entanglement between the superconducting loop and the BEC.

\section*{Acknowledgements}
 The author is very thankful to Prof Peter Hannaford and Prof Tien Kieu for useful and stimulating discussions.

\end{document}